\documentclass{ws-procs975x65}
\newcommand{\lnG}{\widetilde{\ln}}
\newcommand{\expG}{\widetilde{\exp}}
\newcommand{\lambdaG}{\lambda}
\newcommand{\alphasens}{\alpha^{\mathrm{av}}_{\mathrm{sens}}}
\begin{document}

\title{What entropy at the edge of chaos?
\footnote{\uppercase{T}his work was partially supported by
\uppercase{MIUR} (\uppercase{M}inistero
dell'\uppercase{I}struzione, dell'\uppercase{U}niversit\`a e della
\uppercase{R}icerca) under \uppercase{MIUR-PRIN}-2003 project
``\uppercase{T}heoretical Physics of the Nucleus and the Many-Body
Systems''.}}

\author{Marcello Lissia, Massimo Coraddu and Roberto Tonelli}

\address{Ist. Naz. Fisica Nucleare (I.N.F.N.),
Dipart. di Fisica dell'Universit\`a di Cagliari, INFM-SLACS
Laboratory, I-09042 Monserrato (CA), Italy \\
E-mail: marcello.lissia@ca.infn.it}

\maketitle

\abstracts{Numerical experiments support the interesting
conjecture that statistical methods be applicable not only to
fully-chaotic systems, but also at the edge of chaos by using
Tsallis' generalizations of the standard exponential and entropy.
In particular, the entropy increases linearly and the sensitivity
to initial conditions grows as a generalized exponential. We show
that this conjecture has actually a broader validity by using a
large class of deformed entropies and exponentials and the
logistic map as test cases. }

Chaotic systems at the edge of chaos constitute natural
experimental laboratories for extensions of Boltzmann-Gibbs
statistical mechanics. The concept of generalized exponential
could unify power-law and exponential sensitivity to initial
conditions leading to the definition of generalized Liapounov
exponents~\cite{Tsallis:1997}: the sensitivity $ \xi\equiv
\lim_{t\to\infty}\lim_{\Delta x(0)\to 0} \Delta x(t) / \Delta x(0)
\sim  \expG(\lambdaG t)$, where the generalized exponential
$\expG(x) = \exp_q(x) = [1+(1-q)x]^{1/(1-q)}$; the exponential
behavior for the fully-chaotic regime is recovered for $q\to1$:
$\lim_{q\to1} \exp_q(\lambda_q t) = \exp(\lambda t)$. Analogously,
a generalization of the Kolmogorov entropy should describe the
relevant rate of loss of information. A general discussion of the
relation between the Kolmogorov-Sinai entropy rate and the
statistical entropy of fully-chaotic systems can be found in
Ref.~\refcite{Latora:1999prl}: asymptotically and for ideal coarse
graining, the entropy grows linearly with time. The generalized
entropy proposed by Tsallis~\cite{Tsallis:1987eu} $S_q =
(1-\sum_{i=1}^{N} p_i^q)/(q-1)$, with $p_i$ the fraction of the
ensemble found in the $i$-th cell, reproduces this picture at the
edge of chaos; it grows linearly for a specific value of the
entropic parameter $q=q_{\mathrm{sens}}=0.2445$ in the logistic
map: $\lim_{t\to\infty}\lim_{L\to 0} S_q(t)/ t = K_q$. The same
exponent describes the asymptotic power-law sensitivity to initial
conditions \cite{Latora:1999vk}. This conjecture includes an
extension of the Pesin identity $K_q = \lambda_q$. Numerical
evidences with the entropic form
 $S_q $ exist for the
logistic~\cite{Tsallis:1997} and generalized logistic-like
maps~\cite{Tsallis:1997cl}.

Renormalization group methods~\cite{Baldovin:2002ab} yield the
asymptotic exponent of the sensitivity to initial conditions in
the logistic and generalized logistic maps for specific initial
conditions on the attractor;  the Pesin identity for Tsallis'
entropy has been also studied~\cite{Baldovin:2004}.

Sensitivity and entropy production have been studied in
one-dimensional dissipative maps using ensemble-averaged initial
conditions~\cite{Ananos:2004a} and for two simpletic standard
maps~\cite{Ananos:2004b}: the statistical picture has been
confirmed with a different
$q=q_{\mathrm{sens}}^{\mathrm{ave}}=0.35$~\cite{Ananos:2004a}. The
ensemble-averaged initial conditions is relevant for the relation
between ergodicity and chaos and for practical experiments.

 The present study demonstrates the broader applicability of the
 above-described picture by using
the consistent statistical mechanics arising from the
two-parameter
family~\cite{MittalTaneja,Borges1,Kaniadakis:2004nx,Kaniadakis:2004ri}
of logarithms
\begin{equation}\label{eq:logGen}
    \lnG(\xi) \equiv \frac{\xi^{\alpha}-\xi^{-\beta}}{\alpha+\beta}
    \quad .
\end{equation}\
Physical requirements~\cite{Naudts1} on the resulting entropy
select~\cite{Kaniadakis:2004td} $0 \leq \alpha\leq 1 $ and $0\leq
\beta <1 $. \emph{All the entropies} of this class: (i) are
\emph{concave}~\cite{Kaniadakis:2004nx}, (ii) are \emph{Lesche
stable}~\cite{Scarfone:2004ls}, and (iii) yield \emph{normalizable
distributions}~\cite{Kaniadakis:2004td}; in addition, we shall
show that they (iv) yield a \emph{finite non-zero asymptotic rate
of entropy production} for the logistic map with the appropriate
choice of $\alpha$.

We have considered the whole class, but we shall here report
results for three interesting one-parameter cases:

(1) the original Tsallis proposal~\cite{Tsallis:1987eu}
($\alpha=1-q$, $\beta=0$):
\begin{equation}\label{eq:logTsallis}
    \lnG(\xi) =\ln_{q}(\xi) \equiv \frac{\xi^{1-q}-1}{1-q}\quad ;
\end{equation}

(2) Abe's logarithm
\begin{equation}\label{eq:logAbe}
   \lnG(\xi) =  \ln_{A}(\xi) \equiv \frac{\xi^{1/q_A-1}-\xi^{q_A-1}}{1/q_A-q_A}\quad
    ,
\end{equation}
where $q_A=1/(1+\alpha)$ and $\beta=\alpha/(1+\alpha)$, which has
the same quantum-group symmetry of and is related to the entropy
introduced in Ref.~\refcite{Abe:1997qg};

(3) and Kaniadakis' logarithm, $\alpha=\beta=\kappa$, which shares
the same symmetry group of the relativistic momentum
transformation~\cite{Kaniadakis:20012002}
\begin{equation}\label{eq:logKaniadakis}
   \lnG(\xi) = \ln_{\kappa}(\xi) \equiv \frac{\xi^{\kappa}-\xi^{-\kappa}}{2\kappa}\quad
    .
\end{equation}

The sensitivity to initial conditions and the entropy production
has been studied in the logistic map $x_{i+1}=1-\mu x^2_i$ at the
infinite-bifurcation point $\mu_{\infty}=1.401155189$. The
generalized logarithm $\lnG(\xi)$ of the sensitivity, $\xi(t)=
(2\mu)^{t}\prod_{i=0}^{t-1}|x_{i}|$ for $1 \leq t \leq 80 $, has
been uniformly averaged by randomly choosing $4\times 10^{7}$
initial conditions $-1<x_0<1$. Analogously to the chaotic regime,
the deformed logarithm of $\xi$ should yield a straight line
$\lnG(\xi(t))=\lnG(\expG(\lambdaG t))=\lambdaG t$.

Following  Ref.~\refcite{Ananos:2004a}, where the exponent
obtained with this averaging procedure, indicated by
$\langle\cdots\rangle$, was denoted
$q_{\mathrm{sens}}^{\mathrm{av}}$ for Tsallis' entropy, each of
the generalized logarithms, $\langle\lnG(\xi(t))\rangle$, has been
fitted to a quadratic function for $1\leq t \leq 80$ and $\alpha$
has been chosen such that the coefficient of the quadratic term be
zero:  we call this value $\alphasens$.

Statistical errors, estimated by repeating the whole procedure
with sub-samples of the $4\times 10^{7}$ initial conditions, and
systematic uncertainties, estimated by including different numbers
of points in the fit, have been quadratically combined.

We find that the asymptotic exponent $\alphasens=0.650\pm0.005$ is
consistent with the value of Ref.~\refcite{Ananos:2004a}:
$q^{\mathrm{av}}_{\mathrm{sens}}=1-\alphasens\approx 0.36$. The
error on $\alphasens$ is dominated by the systematic one (choice
of the number of points) due to the inclusion of small values of
$\xi$ which produces 1\% discrepancies from the common asymptotic
behavior.

\begin{figure}[htb]
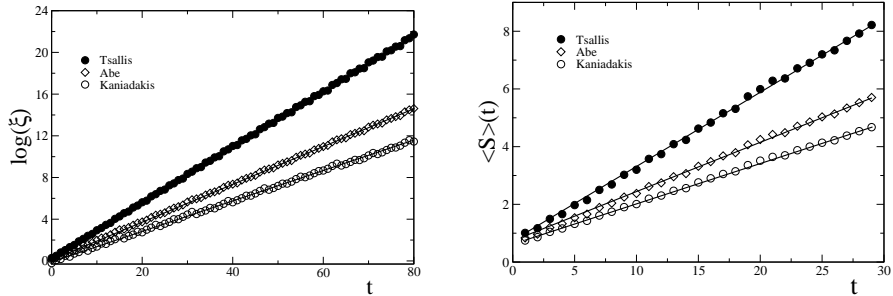

\begin{center}
\includegraphics[width=5.5cm]{sens_xmgr_proceed.eps}\hspace*{18pt}
\includegraphics[width=5.5cm]{ent_xmgr_proceed.eps}
\caption{Generalized logarithm of the sensitivity to initial
conditions (left) and generalized entropy (right) as function of
time. From top to bottom, Tsallis', Abe's and Kaniadakis'
logarithms (entropies) for $\alpha=\alphasens$. In the left frame
the slopes $\lambda$ (generalized Lyapunov exponents) are
$0.271\pm0.004$, $0.185\pm0.004$  and $0.148\pm0.004$; in the
right frame the slopes $K$ (generalized Kolmogorov entropies) are
$0.267\pm0.004$, $0.186\pm0.004$  and $0.152\pm0.004$.
\label{fig:sensitivityAndEntropy}}
\end{center}
\end{figure}

Figure~\ref{fig:sensitivityAndEntropy} shows the straight-line
behavior of $\lnG(\xi)$ for all formulations when
$\alpha=\alphasens$ (right frame); the corresponding slopes
$\lambda$ (generalized Lyapunov exponents) are $0.271\pm0.004$
(Tsallis), $0.185\pm0.004$ (Abe) and $0.148\pm0.004$ (Kaniadakis).
While $\alpha$ is a  universal characteristic of  the map, the
slope $\lambda$ strongly depends on the choice of the logarithm.

The entropy has been calculated by dividing the interval $(-1,1)$
in $W=10^{5}$ equal-size boxes, putting at the initial time
$N=10^{6}$ copies of the system with a uniform random distribution
within one box, and then letting the systems evolve according to
the map. At each time $p_i(t)\equiv n_i(t)/N$, where $n_i(t)$ is
the number of systems found in the box $i$ at time $t$, the
entropy of the ensemble is
\begin{equation}\label{eq:entropyGen}
    S(t) \equiv \left\langle \sum_{i=1}^{N} p_i(t) \lnG(\frac{1}{p_i(t)})\right\rangle =
    \left\langle \sum_{i=1}^{N} \frac{p_i^{1-\alpha}(t)-p_i^{1+\beta}(t)}{\alpha+\beta}
         \right\rangle
\end{equation}
where $\langle\cdots\rangle$ is an average over $2\times 10^{4}$
experiments, each one starting from one box randomly chosen among
the $N$ boxes. The application of the MaxEnt principle to the
entropy~(\ref{eq:entropyGen}) yields as distribution the deformed
exponential that is the inverse function of the corresponding
logarithm of Eq.~(\ref{eq:logGen}):
$\expG(x)=\lnG^{-1}(x)$~\cite{Kaniadakis:2004td}.

\begin{figure}[tb]
\begin{center}
\includegraphics[width=5.5cm]{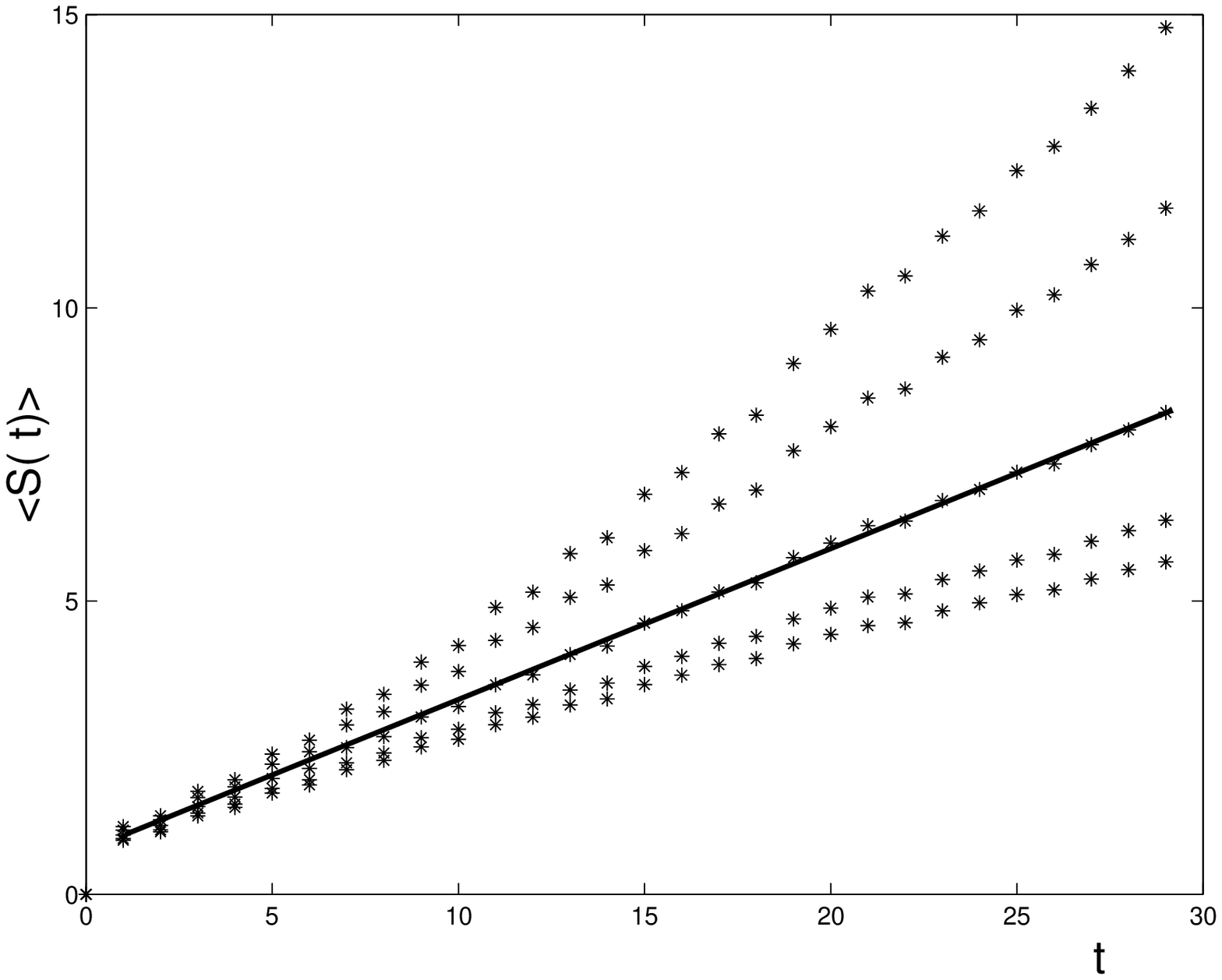}\hspace*{18pt}
\includegraphics[width=5.5cm]{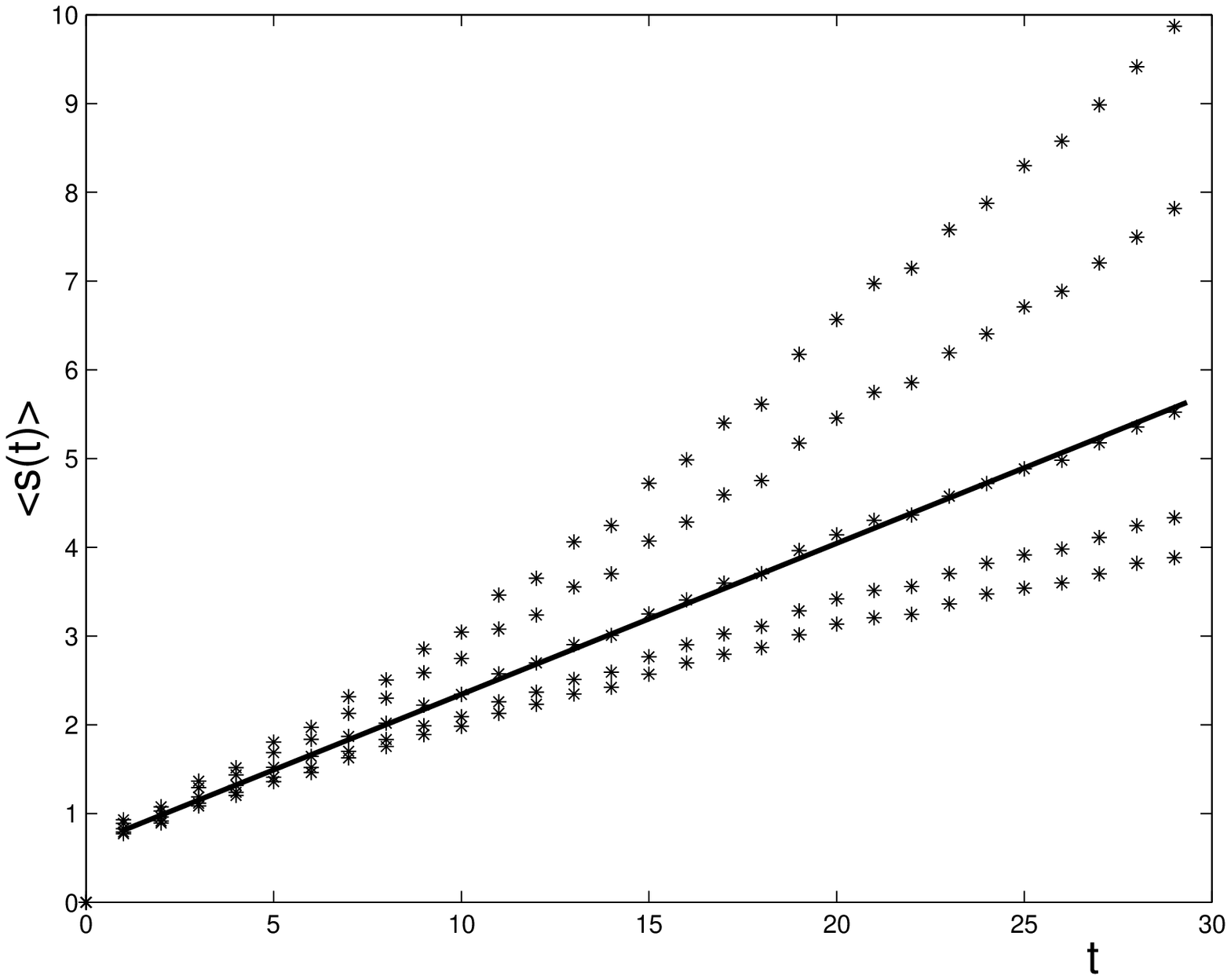}\vspace*{18pt}
\includegraphics[width=5.5cm]{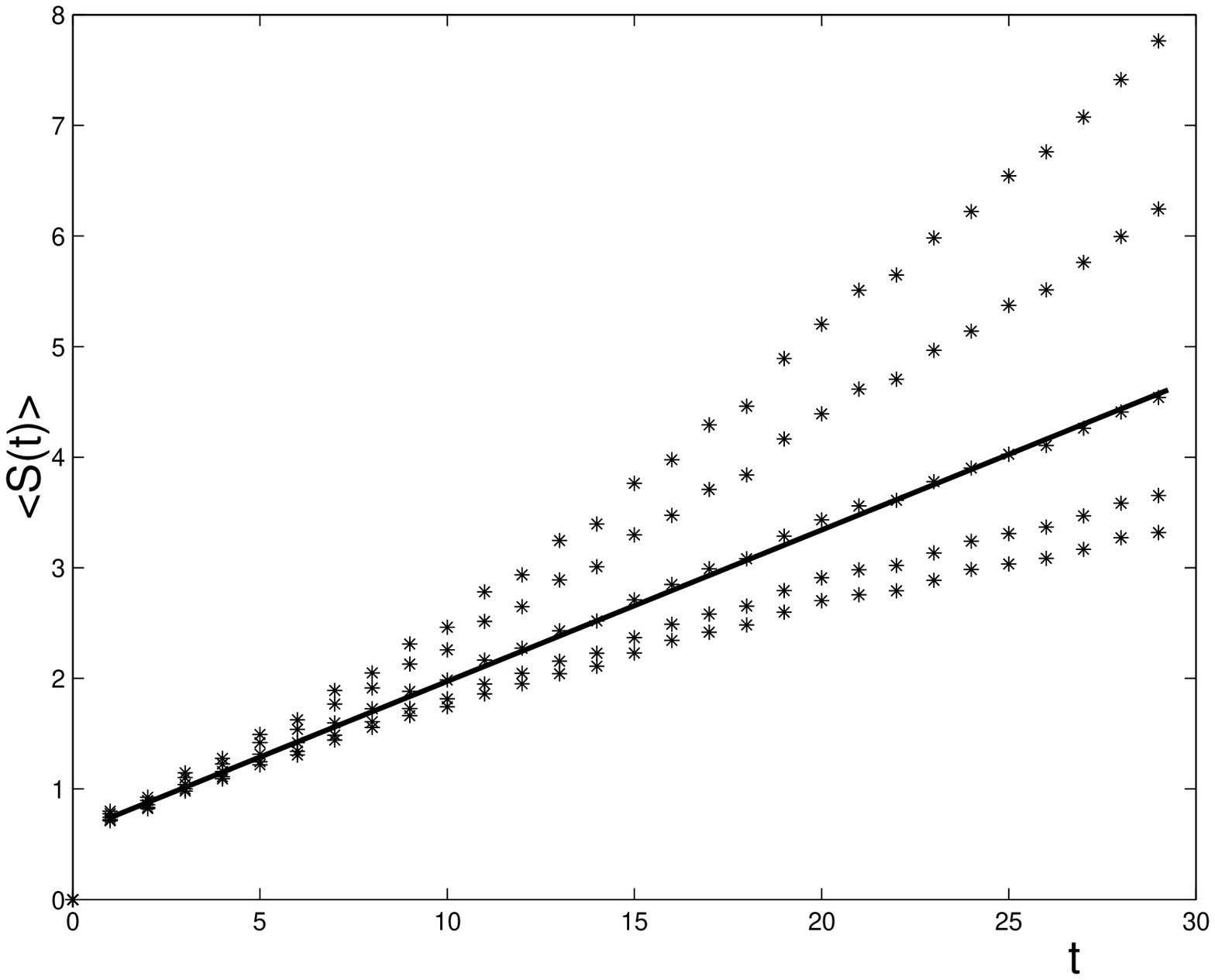}\hspace*{18pt}
 \caption{Tsallis' (top left), Abe's (top right) and Kaniadakis'
 (bottom) entropies as function of time for
 $\alpha=1-q=0.80$, 0.74, 0.64, 0.56, 0.52 (from top to bottom).
 Straight lines are guides for the eyes when $\alpha=0.64\approx
 \alphasens$.
\label{fig:entropyOtherAlphas}}
\end{center}
\end{figure}

Analogously to the strong chaotic case, where an exponential
sensibility ($\alpha=\beta=0$) is associated to a linear rising
Shannon entropy, which is defined in terms of the usual  logarithm
($\alpha=\beta=0$), we use the same values $\alpha$ and $\beta$ of
the sensitivity for the entropy of Eq.~(\ref{eq:entropyGen}).
Fig.~\ref{fig:sensitivityAndEntropy} shows (right frame) that this
choice leads to entropies that grow linearly: the corresponding
slopes $K$ (generalized Kolmogorov entropies) are $0.267\pm0.004$
(Tsallis), $0.186\pm0.004$ (Abe) and $0.152\pm0.004$ (Kaniadakis).
This linear behavior disappears when  $\alpha\neq\alphasens$ as
shown in Fig.~\ref{fig:entropyOtherAlphas} for Tsallis', Abe's and
Kaniadakis' entropies.

In addition, the whole class of entropies and logarithms  verifies
the Pesin identity $K=\lambda$ confirming what was already known
for Tsallis' formulation~\cite{Tsallis:1997,Ananos:2004a}. The
values of $\lambda$ and $K$ for the specific Tsallis', Abe's and
Kaniadakis formulations are given in the caption to
Fig.~\ref{fig:sensitivityAndEntropy} as important explicit
examples of this identity.

 An intuitive explanation of the dependence of the value of $K$
on $\beta$ and details on the calculations can be found in
Ref.~\refcite{Tonelli:2004ha}.

In summary, numerical evidence corroborates and extends Tsallis'
conjecture that, analogously to strongly chaotic systems,  also
weak chaotic systems can be described by an appropriate
statistical formalism. In addition to sharing the same asymptotic
power-law behavior to correctly describe chaotic systems, extended
formalisms should verify precise theoretical requirements.

These requirements define a large class of entropies; within this
class we use the two-parameter formula~(\ref{eq:entropyGen}),
which includes Tsallis's seminal proposal. Its simple power-law
form describes both small and large probability behaviors.
Specifically, the logistic map shows:

(a) a power-low sensitivity to initial conditions with a specific
exponent $\xi\sim t^{1/\alpha}$, where $\alpha = 0.650\pm 0.005$;
this sensitivity  can be described by deformed exponentials with
the same asymptotic behavior $\xi(t)=\expG(\lambdaG t)$ (see
Fig.~\ref{fig:sensitivityAndEntropy}, left frame);

(b) a constant asymptotic entropy production rate (see
Fig.~\ref{fig:sensitivityAndEntropy}, right frame) for trace-form
entropies that go as $p^{1-\alpha}$ in the limit of small
probabilities $p$, where $\alpha$  is the same exponent of the
sensitivity;

(c) the asymptotic exponent $\alpha$ is related to parameters of
known entropies: $\alpha = 1-q$, where $q$ is the entropic index
of Tsallis' thermodynamics~\cite{Tsallis:1987eu};
$\alpha=1/q_A-1$, where $q_A$ appears in the
generalization~(\ref{eq:logAbe}) of Abe's
entropy~\cite{Abe:1997qg}; $\alpha =\kappa$, where $\kappa$ is the
parameter in Kaniadakis' statistics~\cite{Kaniadakis:20012002};

(d) Pesin identity holds $S_\beta /t\to K_\beta = \lambda_\beta$
for each choice of entropy and corresponding exponential in the
class, even if the value of $K_\beta=\lambda_\beta$ depends on the
specific entropy and it is not characteristic of the map as it is
$\alpha$~\cite{Tonelli:2004ha};

\begin{figure}[hp]
\begin{center}
\includegraphics[width=6.5cm]{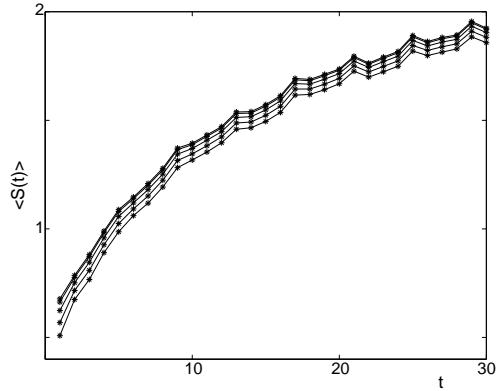}
 \caption{Renyi's entropy for $ 0.1\leq\alpha=1-q\leq 0.95$ (from top to bottom).
 \label{fig:entropyRenyi}}
 \end{center}
\end{figure}

(e) this picture is \emph{not valid} for every entropy: an
important counterexample is the Renyi entropy~\footnote{A
comparison of Tsallis' and Renyi's entropies for the logistic map
can also be found in Ref.~\refcite{Johal:2004}.}, $S_q^{(R)}(t) =
\left\langle (1-q)^{-1}\log\left[\sum_{i=1}^{N}
p_i^q(t)\right]\right\rangle $, which has a non-linear behavior
for any choice of the parameter $q=1-\alpha$ (see
Fig.~\ref{fig:entropyRenyi}).

We gratefully thank S.~Abe, F.~Baldovin, G.~Kaniadakis,
G.~Mezzorani, P.~Quarati, A.~Robledo, A.~M.~Scarfone, U.~Tirnakli,
and C.~Tsallis for suggestions and comments.


\begin{thebibliography}{83}
\bibitem{Tsallis:1997}
C.~Tsallis, A.~R.~Plastino, and W.-M.~Zheng,
{\it Chaos Solitons Fractals} {\bf 8}, 885 (1997).
%
%
\bibitem{Latora:1999prl}
V.~Latora and M.~Baranger,
{\it Phys.\ Rev.\ Lett.\ } {\bf 82}, 520 (1999).
%
\bibitem{Tsallis:1987eu}
C.~Tsallis,
{\it J.\ Statist.\ Phys.\ } {\bf 52}, 479 (1988).
%
\bibitem{Latora:1999vk}
V.~Latora, M.~Baranger, A.~Rapisarda and C.~Tsallis,
{\it Phys.\ Lett.\ A} {\bf 273}, 97 (2000).

\bibitem{Tsallis:1997cl}
U.~M.S.~Costa, M.~L.~Lyra, A.~R.~Plastino and C.~Tsallis,
{\it Phys.\ Rev.\ E }{\bf 56}, 245 (1997).
%

\bibitem{Baldovin:2002ab}
F.~Baldovin and A.~Robledo, {\it Phys.\ Rev.\ E }{\bf 66}, 045104
(2002); {\it Europhys.\ Lett.\ }{\bf 60}, 518 (2002).

\bibitem{Baldovin:2004}
F.~Baldovin and A.~Robledo,
{\it Phys.\ Rev.\ E }{\bf 69}, 045202 (2004).

\bibitem{Ananos:2004a}
Garin~F.~J.~Ananos and Constantino Tsallis,
{\it Phys.\ Rev.\ Lett.\ } {\bf 93}, 020601 (2004).

\bibitem{Ananos:2004b}
G.~F.~J.~Ananos, F.~Baldovin and C.~Tsallis,
arXiv:cond-mat/0403656.
%

\bibitem{MittalTaneja}
D.P. Mittal, Metrika {\bf 22}, 35 (1975); B.D. Sharma, and I.J.
Taneja, Metrika {\bf 22}, 205 (1975).

\bibitem{Borges1} E.P. Borges, and I. Roditi, Phys. Lett. A
{\bf 246}, 399 (1998).


\bibitem{Kaniadakis:2004nx}
G.~Kaniadakis, M.~Lissia, A.~M.~Scarfone,
{\it Physica A} {\bf 340}, 41 (2004).

\bibitem{Kaniadakis:2004ri}
G.~Kaniadakis and M.~Lissia,
Physica A {\bf 340}, xv (2004) [arXiv:cond-mat/0409615].

%
\bibitem{Naudts1} J. Naudts,
Physica A {\bf 316}, 323 (2002) [arXiv:cond-mat/0203489].
%

\bibitem{Kaniadakis:2004td}
G.~Kaniadakis, M.~Lissia, A.~M.~Scarfone,
arXiv:cond-mat/0409683.
%
\bibitem{Scarfone:2004ls}
S. Abe, G. Kaniadakis, and A.M. Scarfone,
{\it J. Phys. A (Math. Gen.)} {\bf 37}, 10513 (2004).


\bibitem{Abe:1997qg}
S.~Abe,
{\it Phys.\ Lett.\ A\ }{\bf 224}, 326 (1997).

\bibitem{Kaniadakis:20012002}
G.~Kaniadakis, {\it Physica A} {\bf 296}, 405 (2001); {\it Phys.\
Rev.\ E\ }{\bf 66}, 056125 (2002).

\bibitem{Tonelli:2004ha}
R.~Tonelli, G.~Mezzorani, F.~Meloni, M.~Lissia and M.~Coraddu,
arXiv:cond-mat/0412730.

\bibitem{Johal:2004}
R.~S.~Johal and U.~Tirnakli,
Physica A {\bf 331}, 487 (2004).
\end{thebibliography}
\end{document}